\documentclass[12pt,oneside,reqno,english]{amsart}
\usepackage{mathptmx}
\usepackage[T1]{fontenc}
\usepackage[latin9]{inputenc}
\usepackage[letterpaper]{geometry}
\usepackage[authoryear]{natbib}

\usepackage{graphicx}
\usepackage{natbib}
\usepackage{hyperref}
\hypersetup{
    colorlinks = true,
    urlcolor   = blue,
    citecolor  = black,
}

\newcommand{\RomanNumeralCaps}[1]
\linenumbers
\DeclareMathOperator{\sech}{sech}
\newcommand{\lyxdot}{.}
\usepackage[letterpaper]{geometry}
\title[Self-induced transparency of long water waves]{Self-induced transparency of long water waves over bathymetry: the dispersive shock mechanism}

\author{Alex Sheremet\textsuperscript{1} and  V. I. Shrira\textsuperscript{2}}
\address{1. Engineering School for Sustainable Infrastructure \& Environment (ESSIE), University of Florida, Gainesville, FL 32611 , USA\\ 2. School of Computing and Mathematics, Keele University, Keele ST5 5BG, UK}

\begin{document}
\maketitle

\begin{abstract}
Dispersive shock waves (DSW) are a salient feature of long water waves
often observed in tidal bores and tsunami/meteotsunami contexts. Their
interaction with bathymetry is poorly understood. The shoreline hazard
from tsunamis and meteotsunamis critically depends on the fraction
of incoming energy flux transmitted across the shallow nearshore shelf.
Here, by considering nonlinear dynamics of waves over variable depth
within the framework of the Boussinesq equations we show that the
transmitted energy flux fraction can strongly depend on the initial
amplitude of the incoming wave and the distance it travels. The phenomenon
is similar to self-induced transparency in nonlinear optics: small
amplitude waves are reflected by bathymetry inhomogeneity, while a
larger amplitude ones pass through. The mechanism of self-induced
transparency of long water waves can be explained as follows. In linear
setting a bathymetry inhomogeneity, of length comparable to that of
the incident wavelength, by transmitting high wavenumber components acts
as a high-pass filter. The DSW evolution efficiently transfers wave
energy into high wavenumber band, where reflection is negligible.
By examining an idealized model of bathymetry we show that this is
an order one effect and explore its dependence on parameters in the
range relevant for meteotsunamis. The role of wave energy transfer
into high wavenumber band owing to the growth of bound harmonics unrelated
to the DSW was found to be small. 
\end{abstract}



\section{Introduction\label{sec: intro}}

Nonlinear optics and plasmas exhibit ``self-induced transparency''
\citep{Kocharovskaya1986}, a phenomenon in which quantum interference
allows for the propagation of high intensity light through an otherwise
opaque medium. 

A functionally similar phenomenon can occur for long water waves propagating
over inhomogeneous bathymetry. The essence of the idea is simple.
Linear wave transmission by a localized inhomogeneity acts as a high-pass
filter, long waves are strongly reflected, while short waves pass
through \citep{Meyer1979,Meyer1975,Mei2005,ErmakovStepanyants2020}.
However, if the long wave is nonlinear, a significant amount of energy
is transferred to shorter wave scales, and the reflected/transmitted
energy balance changes.

As it happens, long ocean waves do exhibit a peculiar scale cascade
mechanism. Under the combined effects of nonlinearity and dispersion,
long waves can disintegrate into a train of short pulses. A celebrated
example are the ``undular bores'', observed at tidal fronts in rivers
(e.g., \citealp{Rayleigh1914,Lamb1932,Benjamin1954,Chanson2011};
more recent works by \citealp{El2012,El2005}, and others). ``Undular
bore''-like wave patterns are in fact a universal feature of weakly
nonlinear evolution of weakly dispersive waves, with or without inhomogeneity,
encountered a wide variety of physical contests, such as nonlinear
optics \citep[e.g.,][]{Wan2007,Fatome2014}, flows of Bose-Einstein
condensates \citep{Xu2017}, internal waves in the ocean and the atmosphere
\citep{Porter2002}, and many others. Their formation and evolution
was studied mostly within the Korteweg - de Vries (KdV) framework
(\citealp{Gurevich1974,Karpman1967,Karpman1975,Whitham1973}; and
a recent review by \citealp{Kamchatnov2021}), where the integrability
of the governing equation greatly facilitates the analysis. Consider
an initially smooth localized perturbation of the free surface with
maximum height $a$ of characteristic length $L$ in water of characteristic
depth $h_{0}$, characterized by the nonlinearity and dispersion parameters
$\epsilon=a/h_{0}$ and $\mu=h_{0}^{2}/L^{2}$, both assumed small.
If initially $\mu\ll\epsilon$, the perturbation dynamics may be approximately
described as that of a Riemann wave with no dispersion \citep[e.g.,][]{Whitham1973}.
Because the crest of the wave propagates faster than the trough the
wave steepens, as the wave approaches gradient catastrophe, dispersion
becomes important at the front of the wave, causing it to disintegrate
into much shorter waves for which nonlinearity and dispersion are
in balance. In the KdV framework, these shorter waves are the close
to KdV solitons \citep[e.g.,][]{KadomtsevKarpman1971,Karpman1975,Kamchatnov2012,El2005,El2012}.
Although in different geographical locations and different physical
contexts the phenomenon is known under many different names \citep[e.g.,][]{Chanson2011},
for obvious reasons, the general name for this process is ``dispersive
shock waves'' (DSW); this term will be used throughout the paper.
\begin{figure}\centering
\includegraphics[width=1\textwidth]{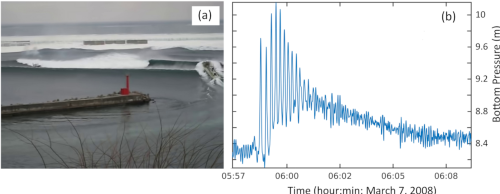}\caption{Examples of tsunami and meteotsunami DSWs. Left: Fukushima tsunami
(2011) at Kuji Port, Iwate Prefecture, Japan; frame 1:20 min from
video posted by \citet{Fukushima2011}. The tug boat length is $\approx$
20 m long (visual estimation). Right: meteotsunami DSW or solibore
recorded in near the 8-m isobath on the Atchafalaya shelf, LA, USA
\citep{Sheremet2016}. \label{fig: mt bore}}
\end{figure}

Here, we show that the disintegration of long ocean wave into a DSW
pattern transfers enough energy flux to short scales to significantly
enhance the transmission of the long wave energy past a bathymetric
inhomogeneity. Because DSW form faster for larger amplitude waves,
the DSW reflection process is truly similar to ``self-induced transparency'':
simply stated, at a bathymetric inhomogeneity, a small amplitude long
wave reflects, while a larger amplitude one goes through. 

Long ocean waves susceptible of DSW self-induced transparency include
tsunamis and meteotsunamis (figure \ref{fig: mt bore}). Tsunamis
are generated in the deep ocean primarily by earthquakes or volcanic
activity. Due to their characteristic dimensions (length $L\sim100$
km, amplitude $a\sim1$ m) tsunamis generated in the deep ocean evolve
over the open ocean effectively as linear waves and may develop a
DSW only on the shelf \citep[e.g.,][]{Madsen2008}. In contrast, meteotsunamis
are shorter waves ($L\sim10$ km, $a\sim0.5$ m), typically generated
on the continental shelf by atmospheric perturbations through Proudman
\citep[e.g.,][]{Proudman1929,Rabinovich1996,Rabinovich1998,Monserrat2006,Vilibic2014,Pellikka2022}.
Assuming a constant shelf slope of $5\times10^{-4}$, table \eqref{tab: mt pars}
shows the characteristic evolution for a 1-m height tsunami entering
the shelf at 100 m depth, and a 0.5-m amplitude meteotsunami generated
at near the 50-m isobath. The location of the gradient catastrophe,
estimated using the nonlinear shallow water equations with no dispersion,
is similar fot the two waves, at depth $h_{c}$ in the order of 10
m. Note that the DWS disintegration occurs before the gradient catastrophe.
\begin{table}
\begin{centering}
{\footnotesize{}}%
\begin{tabular}{cccccccc}
{\footnotesize{}$T$ (min)} & {\footnotesize{}$h$ (m)} & {\footnotesize{}$L$ (km)} & {\footnotesize{}$a$ (m)} & {\footnotesize{}$\epsilon$} & {\footnotesize{}$\mu$} & {\footnotesize{}$\sigma^{2}$} & {\footnotesize{}$h_{c}$ (m)}\tabularnewline
{\footnotesize{}15} & {\footnotesize{}$100$} & {\footnotesize{}28} & {\footnotesize{}1} & {\footnotesize{}0.01} & {\footnotesize{}$1.3\times10^{-5}$} & {\footnotesize{}794} & {\footnotesize{}8.2}\tabularnewline
{\footnotesize{}4} & {\footnotesize{}50} & {\footnotesize{}6.6} & {\footnotesize{}0.5} & {\footnotesize{}$0.01$} & {\footnotesize{}$5\times10^{-5}$} & {\footnotesize{}176} & {\footnotesize{}9.2}\tabularnewline
\end{tabular}{\footnotesize\par}
\par\end{centering}
\caption{Characteristic scales of tsunamis and meteotsunamis on the shelf of
slope $\approx5\times10^{-4}$ \citep{Madsen2008,Sheremet2016}; $T$,
$L$, and $a$ are the wave time, length and amplitude scales; $\epsilon$
and $\mu$ are the nonlinearity and dispersion parameters; $h$ is
the depth at wave origin (shelf edge for tsunamis); $h_{c}$ is the
depth at the location of gradient catastrophe estimated ignoring dispersion.
\label{tab: mt pars}}
\end{table}

The shoreline hazard posed by long waves such as tsunamis and meteotsunamis
depends critically on the processes affecting its propagation from
the deep ocean onto the shallow nearshore shelf. While DSW solitons
are a hazard in its own right, as illustrated in figure (\ref{fig: mt bore}a)
by the precarious pitch of the tugboat attempting to evade them, the
increased transmission of energy due to self-induced transparency
may play a significant role in the shoreline impact of the wave. 

To the best of our knowledge, the DSW self-induced transparency process
has not been studied. Here, we investigate the conditions favorable
for this phenomenon to occur and provide its quantitative description
within the framework of a simplified model. In \S\ref{sec: problem}
we formulate the problem and present and discuss mathematical and
the numerical tools we use to investigate it. The relevant aspects
of the DSW evolution over even bottom are discussed in \S\ref{subsec: DSW}.
The results of the analysis are presented in \S\ref{sec: results},
while \S\ref{sec: discussion} concludes with the discussion of the
phenomenon, its modeling and the underlying assumptions.

\section{Formulation of the problem \label{sec: problem}}

\subsection{Basic equations, assumptions and simplifications}

We consider a classical problem of the nonlinear evolution of long
water waves over inhomogeneous bathymetry assuming the water depth
to be small but finite compared to characteristic wave length. The
viscous effects and wave interaction with atmosphere are neglected.
A suitable formal framework for this problem is the Boussinesq approximation
(e.g., \citealp{Peregrine1967,Whitham1973,Karpman1975,Mei2005}; see
also \citealp{Dingemans1997}); the choice of a particular version
of the Boussinesq equations is immaterial. For simplicity only we
consider one-dimensional geometry 
\begin{equation}
\eta_{t}+[\left(h+\eta\right)u]_{x}=0,\quad u_{t}+g\eta_{x}+uu_{x}=\frac{1}{2}h\left[\left(hu\right)_{xx}-\frac{1}{3}hu_{xx}\right]_{t},\label{eq: Bouss}
\end{equation}
where $\eta(x,t)$ is the free surface elevation, $x$ is the horizontal
coordinate, $t$ is time, $u(x,t)$ is the horizontal flow velocity,
$h(x)$ is the bathymetry, $g$ is the gravity acceleration, and the
subscripts $x$ and $t$ denote partial derivatives. 

The Boussinesq equations describe both left and right propagating
waves as well as their interactions, and therefore provide a full
description of the scattering process. However, the interplay of nonlinearity,
dispersion and inhomogeneity in the wave evolution is usually too
complex to be described analytically, and even with numerical models,
it is difficult to deconstruct it into its basic mechanisms. Alternatively,
if physics allows for considering only left- or right-propagating
waves alone, equations \eqref{eq: Bouss} simplify to the Korteweg-Vries
(KdV) equation with variable coefficients \citep{Ostrovsky1975}.
KdV simplifications provided much of the analytical understanding
of the DSW process \citep[e.g.,][]{Gurevich1974,Karpman1975,El2005,El2012,EL2016,Kamchatnov2021},
but it is ostensibly inapplicable to the scattering problem. 

To deconstruct the interplay between nonlinearity, dispersion, and
inhomogeneity, we consider in some detail a strongly idealized bathymetric
model.

\subsection{Model bathymetry}

To make the problem maximally tractable and transparent we separate
in space the domains of wave nonlinear evolution and interaction with
bathymetry. Remarkably, such a separation is indeed possible as a
rough approximation in many real situations. We exploit the observation
that in the ocean very often the areas of the noticeable bathymetry
inhomogeneity, e.g. continental shelf breaks, separated by the relatively
flat areas of the abyssal plane and the shelf. This bathymetry pattern
also repeats itself for smaller scales closer to the shoreline. For
brevity only we will refer to the flat areas as ``shelves'' and areas
of steep inhomogeneity as ``slopes'' throughout the paper, irrespective
of the scales.

\begin{figure}\centering
\includegraphics[width=0.8\textwidth]{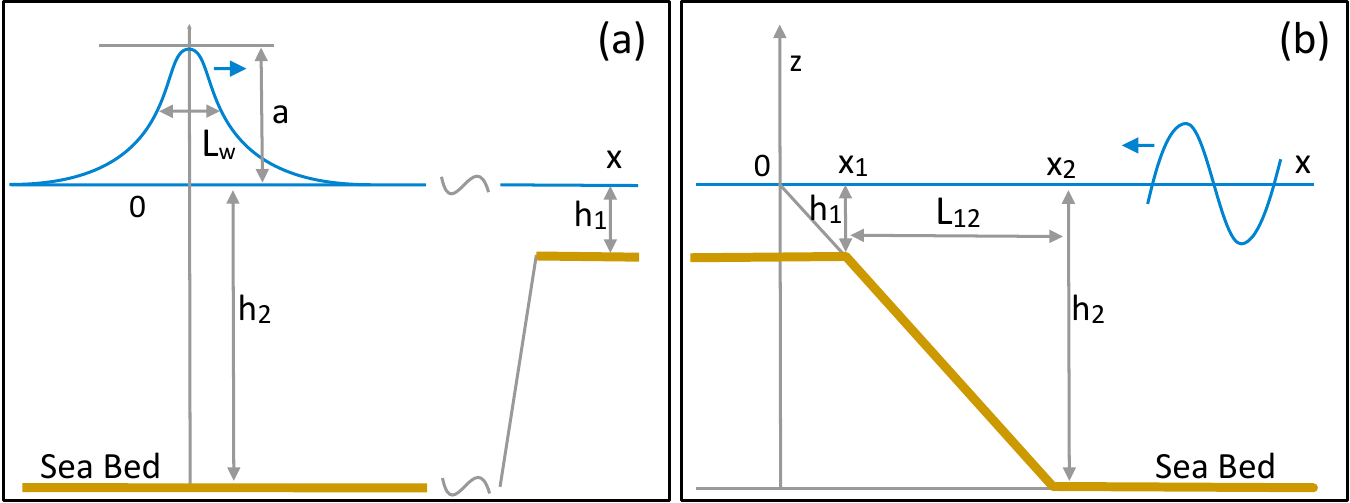} \caption{a) Simplified reflection/transmission problem for a weakly nonlinear
positive perturbation. The perturbation propagates over constant depth
and encounters a slope. Varying the distance from the initial position
of the perturbation to the slope toe allows for investigating the
reflection/transmission of the waves at various stages of their evolution.
b) Bathymetry settings for reflection/transmission coefficients \citep{ErmakovStepanyants2020}.
\label{fig: problem} }
\end{figure}

Consider a long initial perturbation propagating over a domain of
constant depth $h_{2}$ of arbitrary horizontal extent towards a segment
of constant slope where the depth changes to $h_{1}\ne h_{2}$ (figure
\ref{fig: problem}a). Assuming that the slope to be steep enough,
so that the long-wave evolution on the slope is fast compared to the
timescale of nonlinearity, enables us to treat wave dynamics on the
slope and wave reflection by the slope as linear (see a brief discussion
below in \S\ref{sec: discussion}). By varying the distance between
the initial position of the perturbation and the slope toe, it is
easy to examine the effect of reflection at different stages of wave
nonlinear evolution. This model does not decouple reflection from
nonlinear evolution, it decouples concurrent reflection from nonlinear
evolution. Crucially, it also greatly simplifies the nonlinear evolution
and reflection calculations, because the evolution on the flat domain
of constant depth $h_{2}$ the evolution of the incoming perturbation
may be calculated using the unidirectional KdV simplification of the
Boussinesq equations \eqref{eq: Bouss}. 

\subsection{Reflection and transmission at a constant slope\label{subsec: RT}}

Since it is known that in the linear setting an inhomogeneity in topography
is acts as a high-pass filter, reflecting longer wave while allowing
the shorter components to pass through, we confine our discussion
to reflection and transmission of longer waves components. This is
equivalent to neglecting the right-hand side of equations \eqref{eq: Bouss},
i.e., reducing them to the classical non-dispersivs shallow water
equations. A comprehensive discussion of linear reflection of nondispersive
waves by several analytic slope shapes is given in \citet{ErmakovStepanyants2020}
(see also references therein). Here, we provide a brief overview of
the relevant results. We confine this discussion to the simplest shape
of the slope --the constant slope, but the results discussed here
are of general applicability.

On the slope $\alpha=\frac{h_{2}-h_{1}}{L_{12}}$ (figure \ref{fig: problem}b),
the linearized shallow water equations read 
\begin{equation}
\eta_{t}+u_{x}h+uh_{x}=0,\quad u_{t}+g\eta_{x}=0.\label{eq: swe lin}
\end{equation}
Since the equations are linear we perform the Fourier transform with
respect to time and consider monochromatic in time solutions. The
resulting system then reduces to the Bessel equation with the known
solution in terms of the Bessel functions of the first and second
kind $J_{n}$ and $Y_{n}$, 
\begin{align}
\eta & =\alpha L_{12}\frac{i}{\varpi}\left[AJ_{0}\left(2\varpi\sqrt{\xi}\right)+BY_{0}\left(2\varpi\sqrt{\xi}\right)\right]e^{i\varpi\tau}.\label{eq: eta}\\
u & =\frac{L_{12}}{T_{s}}\frac{1}{\varpi\sqrt{\xi}}\left[AJ_{1}\left(2\varpi\sqrt{\xi}\right)+BY_{1}\left(2\varpi\sqrt{\xi}\right)\right]e^{i\varpi\tau}.\label{eq: u}
\end{align}
where $A$ and $B$ are arbitrary constants, the variables are scaled
using the length of the slope $L_{12}$ (see figure \ref{fig: problem}b)
$x=L_{12}\xi$, $h=h_{1}-\alpha L_{12}\xi$, and $\varpi=\omega T_{12}$,
where $\omega$ is the angular frequency, and 
\begin{equation}
T_{12}=\sqrt{\frac{L_{12}}{\alpha g}}\label{eq: T12}
\end{equation}
is the ``slope'' time scale.

The complete solution for the reflection of a single harmonic of frequency
$\omega$ is obtained by matching at $x_{1,2}$ the slope solution
(\eqref{eq: eta}-\eqref{eq: u}) with monochromatic waves propagating
to in both directions of the two shelves. This enables one to find
the reflection and transmission coefficients 
\begin{equation}
\mathcal{R}_{\uparrow}=\frac{z_{J1}^{*}z_{Y2}^{*}-z_{J2}^{*}z_{Y1}^{*}}{z_{J1}^{*}z_{Y2}-z_{J2}z_{Y1}^{*}};\quad\mathcal{T}_{\uparrow}=\frac{z_{J1}^{*}z_{Y1}-z_{J1}z_{Y1}^{*}}{z_{J1}^{*}z_{Y2}-z_{J2}z_{Y1}^{*}},\label{eq: RT up}
\end{equation}
for a wave propagating upslope, and 
\begin{equation}
\mathcal{R}_{\downarrow}=\frac{z_{J1}z_{Y2}-z_{J2}z_{Y1}}{z_{J1}^{*}z_{Y2}-z_{J2}z_{Y1}^{*}};\quad\mathcal{T}_{\downarrow}=\frac{z_{J2}^{*}z_{Y2}-z_{J2}z_{Y2}^{*}}{z_{J1}^{*}z_{Y2}-z_{J2}z_{Y1}^{*}}\label{eq: RT down}
\end{equation}
for a wave propagating downslope, where 
\begin{equation}
z_{Jn}=J_{0}\left(2\varpi\sqrt{\xi}_{n}\right)+iJ_{1}\left(2\varpi\sqrt{\xi}_{n}\right);\;z_{Yn}=Y_{0}\left(2\varpi\sqrt{\xi}_{n}\right)+iY_{1}\left(2\varpi\sqrt{\xi}_{n}\right).\label{eq: zJYn}
\end{equation}
and asterisks denote complex conjugates.

\begin{figure}\centering
\includegraphics[width=0.5\textwidth]{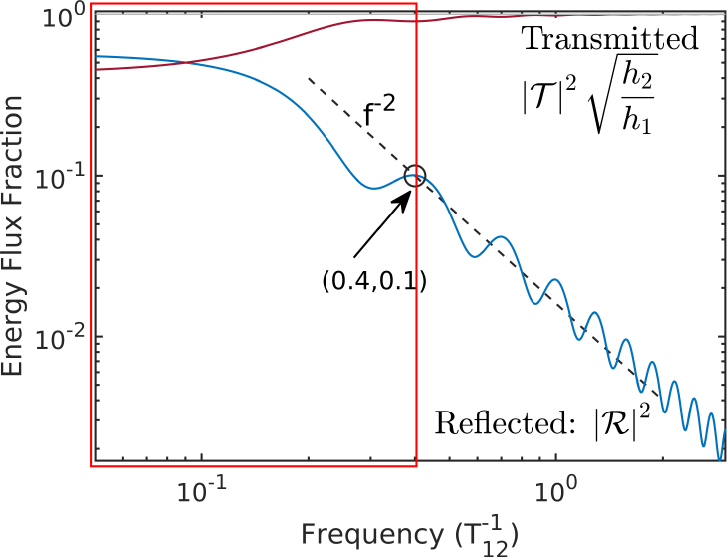}\caption{Reflected and transmitted fraction of energy flux for monochromatic
wave by a constant slope for $h_{2}/h_{1}=50$ (log scale; see also
figure \ref{fig: problem}b). The reflected fraction is below 10\%
for all frequencies greater than $0.4\ T_{12}^{-1}$ (equation \ref{eq: T12}).
The reflection coefficient decays roughly as $f^{-2}$ outside the
red box. \label{fig: RT}}
\end{figure}

The moduli of the coefficients do not depend on the direction of propagation:
$\left|\mathcal{R}_{\uparrow}\right|=\left|\mathcal{R}_{\downarrow}\right|$
and $\left|\mathcal{T}_{\uparrow}\right|=\left|\mathcal{T}_{\downarrow}\right|$.
The reflection and transmission coefficients given by equation (\ref{eq: RT up})
satisfy the conservation of energy, in the sense that the energy flux
of the incoming wave is equal to the sum of the energy fluxes of the
reflected and transmitted waves, $\left|\mathcal{R}\right|^{2}+\left|\mathcal{T}\right|^{2}\sqrt{\frac{h_{1}}{h_{2}}}=1$
(direction subscript discarded). In this relation, $\left|\mathcal{R}\right|^{2}$
may be interpreted as the fraction of the energy flux reflected by
the slope, and $\left|\mathcal{T}\right|^{2}\sqrt{\frac{h_{1}}{h_{2}}}$
represents the fraction of the incoming energy flux transmitted past
the slope. Figure (\ref{fig: RT}) shows the reflected and transmitted
energy flux fractions for $\frac{h_{2}}{h_{1}}=50$ (e.g., e depth
change from 50 m to 1 m). The reflected and transmitted energy flux
fractions have identical frequency distributions for the same ratio
of depths, In the long wave limit, the reflection and transmission
coefficients take the well known forms (\citealp{Mei2005}; constant
gain, no phase shift) 
\begin{align}
\mathcal{R}(0) & \sim\frac{\sqrt{\xi_{2}}-\sqrt{\xi_{1}}}{\sqrt{\xi_{1}}+\sqrt{\xi_{2}}},\;\mathcal{T}(0)\sim\frac{2\sqrt{\xi_{2}}}{\sqrt{\xi_{1}}+\sqrt{\xi_{2}}},\label{eq: RT long wave}
\end{align}

The reflected energy fraction falls below 10\% for frequencies exceeding
0.4 $T_{12}^{-1}$ (outside the red box in figure \ref{fig: RT}).
Figure \eqref{fig: RT} illustrates the statement that reflection
and transmission processes may be described as complementary linear
low- and high-pass filters, with the spectral response provided by
the reflection (transmission) coefficients given by equations \ref{eq: RT up}-\ref{eq: RT down}.
This is consistent with our consideration of reflection within the
framework of the shallow water equation, neglecting high frequency
dispersion.

\section{DSW evolution over a constant depth segment\label{subsec: DSW}}

Here we examine wave evolution over the flat bottom segment prior
to its encountering inhomogeneity of bathymetry. To this end, we consider
unidirectional evolution of a weakly nonlinear wave weakly dispersive
perturbation $\eta(x,t)$ over the entire domain of constant depth
$h$. The Boussinesq equations can be reduced to the single KdV) equation
for the unidirectional evolution, even, we stress it, in the presence
of reflected waves propagating in the opposite direction under mild
additional conditions which we briefly discuss in \S\ref{sec: discussion},
\citep{Whitham1973,Karpman1975}. The reason enabling us to neglect
nonlinear interaction with the weakly nonlinear counterpropagating
waves is that we are concerned only with the localized initial perturbations
which pass through each other too quickly for nonlinear effects to
accumulate.

On non-dimensionalizing the variables as specified below and casting
the equation in terms of signaling coordinates which are more convenient
for studying the evolution of a perturbation over a finite domain,
the resulting KdV equation takes the standard form 
\begin{equation}
\varphi_{\xi}+\varphi\varphi_{\tau}+\frac{1}{\sigma^{2}}\varphi_{\tau\tau\tau}=0,\label{eq: tkdv scaled}
\end{equation}
where $\eta=a\varphi$, with $a$ being the amplitude of the initial
perturbation; and $x=\frac{a}{T}\xi$ and $t=T\tau$. The KdV equation
\eqref{eq: tkdv scaled} is non-dimensionalized using the characteristic
time scale $T$ of the initial perturbation. The parameter $\sigma^{2}=\frac{\epsilon}{\mu}$
the Ursell number, $\epsilon=\frac{a}{h}$ and $\mu=\frac{h^{2}}{L^{2}}$,
$L=3cT$ and $c=\sqrt{gh}$. Equation \eqref{eq: tkdv scaled} should
be viewed as the Cauchy problem with the boundary and initial conditions
\begin{equation}
\varphi(0,\tau)=\varphi_{0}(\tau);\quad\varphi(\xi>0,0)=0.\label{eq: kdv IC}
\end{equation}
Equation \eqref{eq: tkdv scaled} has an infinite number of conserved
quantities \citep{Whitham1965,Miura1968,Karpman1975} of the form
\begin{equation}
Q_{m}=\int_{-\infty}^{\infty}q_{m}(\eta)dx,\;m=1,2,\cdots\label{eq: Q}
\end{equation}
where $q_{m}$ are polynomials of order $m$ in $q$ and may contain
its spatial derivatives. The lowest orders $q_{1}=a\varphi$, $q_{2}=a\varphi^{2}/2$
may be interpreted as momentum and energy densities. The equation
is exactly solvable by the Inverse Scattering Transform technique
\citep[e.g.,][]{Whitham1965,Karpman1975,AblowitzSegur1981} and other
methods. Solitons of the form $\varphi\propto\sech^{2}$ play fundamental
role in the solution of the Cauchy problem: solitons emerge as the
large-time asympotics of any initially localized perturbation of positive
polarity. Equation \eqref{eq: tkdv scaled} is scaled in such a way
that the scale of a soliton of equation \eqref{eq: tkdv scaled} correspond
to $\sigma_{s}^{2}=12$. If $\int_{-\infty}^{\infty}\varphi_{0}(\tau)d\tau\ge0$,
the initial pulse disintegrates into a number of solitons and a dispersive
tail. If the disturbance $\varphi_{0}$ is strongly nonlinear, $\sigma^{2}\gg\sigma_{s}^{2}$.
Initially, the wave evolves as a Riemann simple wave without dispersion:
the front steepens, the shock begins to form. However, as the wave
front approaches the gradient catastrophe, dispersive shock is formed:
at the front of the wave dispersion becomes important, causing the
wave to disintegrate into wave groups that eventually transform into
solitons.In our context the most interesting and relevant are the
situations with large $\sigma$ generating $N(N\gg1)$ solitons 
\begin{equation}
N=\frac{\sigma}{\pi\sqrt{6}}\int_{\varphi(\xi)>0)}\sqrt{\varphi(\xi)}d\xi.\label{eq: N}
\end{equation}
Note, that however small is the initial pulse, provided that the integral
$\int_{-\infty}^{\infty}\varphi_{0}(\tau)d\tau$ is positive at least
one soliton can always form (but only if the flat bottom extent is
infinite).

Consider as an example a single hump initial condition for which simple
expressions are known for the amplitudes and number of the emerging
solitons under assumption of the infinite extent of the flat bottom
(e.g. \citep{Karpman1975}) 
\begin{equation}
\varphi(\xi)=a\sech^{2}\frac{\tau}{T}.\label{eq: sech2}
\end{equation}
This initial condition produces no tail, while the solitons have amplitudes
\citep{Karpman1975} 
\begin{equation}
\frac{a_{n}}{a}=\frac{3}{\sigma^{2}}\left[1+\sqrt{1+\frac{2}{3}\sigma^{2}}-2n\right]^{2},\;\text{with}\;n=1,2,\cdots,N<\frac{1}{2}\left(1+\sqrt{1+\frac{2}{3}\sigma^{2}}\right).\label{eq: sol}
\end{equation}
The \emph{n}-th soliton in the sequence carries the values of the
\emph{m}-th conservation integral $Q_{m,n}$ as follows, 
\begin{equation}
Q_{m,n}=\sqrt{\frac{12}{\sigma^{2}}}\frac{2^{m}\left[\left(m-1\right)!\right]^{2}}{2m-1}a_{n}^{\frac{\left(2m-1\right)}{2}}.\label{eq: solQ}
\end{equation}
\begin{figure}\centering
\includegraphics[width=0.5\textwidth]{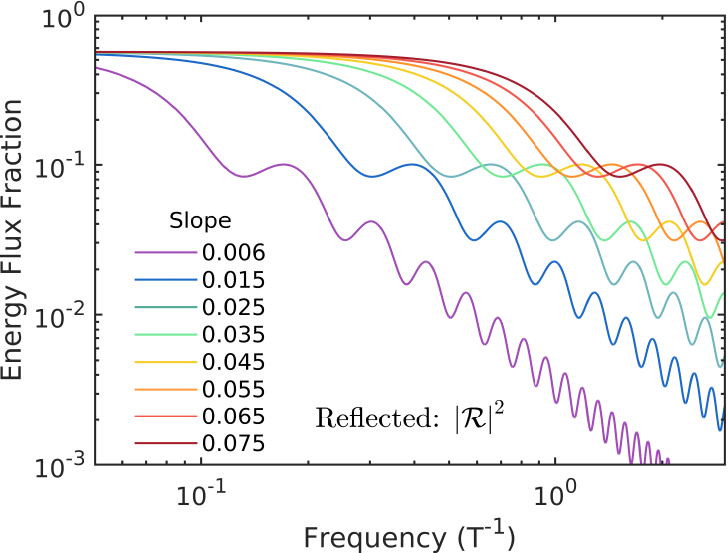}\caption{Dependence of the reflected fraction of energy flux for a monochromatic
wave on the slope (see figure \ref{fig: problem}a). The reflection
coefficient is plotted as function of the time scale (inverse period)
of the wave. A comparison with figure \eqref{fig: RT} suggests that
$T\approx T_{12}$ for slope $\alpha=0.015$. \label{fig: RT-slopes}}
\end{figure}

These theoretical results are valid for infinite
time and infinite extent of the horizontal segment, which is often
far from the realistic situations. Nevertheless, these results enable
us to get important \emph{a priori} bounds on reflected and transmitted
energy which will briefly discuss below in \S\ref{sec: discussion}.

\section{The DSW effect on reflection. Numerical simulations\label{sec: results}}

\subsection{Simulations\label{sec: simulations}}

The asymptotic state of the disintegration of a DSW is well understood
\citep[e.g.,][]{Gurevich1974,Karpman1975,Whitham1973,Caputo2003,El2005,El2012,Kamchatnov2012,Kamchatnov2021},
there are even explicit solutions \citealp{Opanasenko2023}. However,
for a practical application to the reflection of a long wave by an
inhomogeneity, one expects that a significant part of the reflection
process occurs during transient states of DSW disintegration, well
before the wave reaches its asymptotic state. Thus, at present there
is no alternative to resorting to numerical simulation of the wave
evolution. 

\begin{figure}\centering
\includegraphics[width=0.8\textwidth]{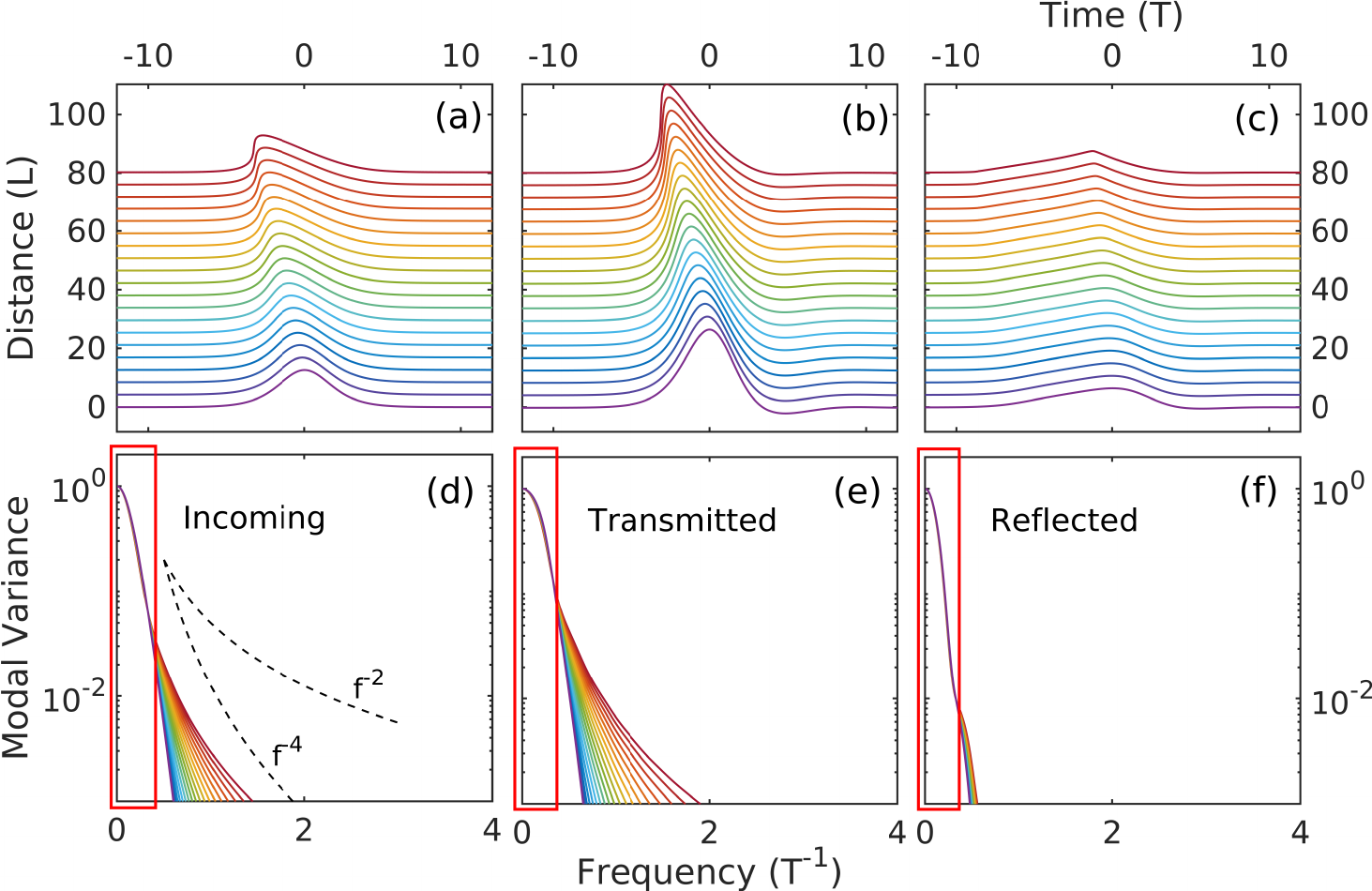}

\caption{Reflection/transmission at different stages of the nondispersive evolution
of a perturbation with $\sigma^{2}=397$ (reference wave amplitude
$a=0.5$ m). Upper panels (a-c): free surface elevation for the incoming
(left), transmitted (middle) and reflected (right) waves. Lower panels
(d-f): Modal variance normalized by the maximum value. Lines represent
wave shapes produced if the slope toe were positioned at the location
indicated. The red box indicates the spectral band subjected to strong
reflection. The reflected energy flux fraction outside the red box
in the lower panels is < 10\%. Dashed lines in panel (d) plot the
frequency dependence of the modal variance for the Fourier series
of a step function ($f^{-2}$), and triangle wave ($f^{-4}$). \label{fig: na0.5}}
\end{figure}

To understand the effect of DSW on wave reflection we have first to
examine the intermediate stages of the DSW evolution, to this end
we integrate the KdV equation \eqref{eq: tkdv scaled} numerically
for an initial disturbance (equation \ref{eq: kdv IC})\textcolor{red}{{}
}of the form 
\begin{equation}
\varphi_{0}(\tau)=a\sech^{2}\frac{\tau-\tau_{0}}{T},\label{eq: perturbation}
\end{equation}
for which the solutions for large time are available in analytical
form, see e.g. \citep{Karpman1967,Karpman1975}. As we discuss below,
these explicit solutions could provide an upper bound on the reflection
flux.

\begin{figure}\centering
\includegraphics[width=0.8\textwidth]{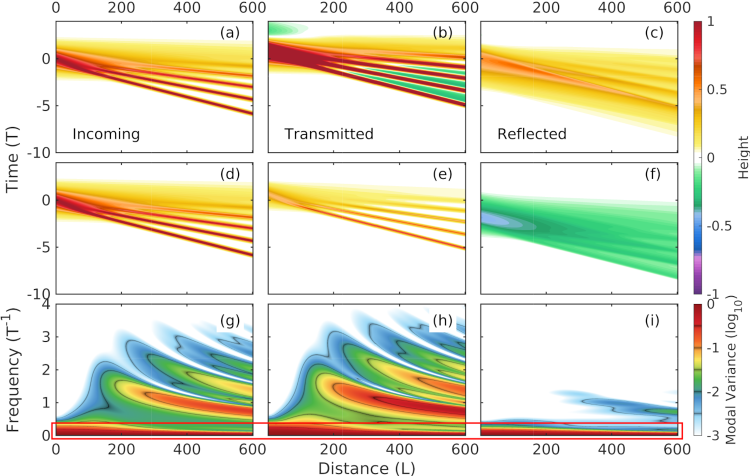}

\caption{Evolution of perturbation with $\sigma^{2}=397$ (reference wave amplitude
0.5 m). Panel columns: left -- incoming wave; middle -- transmitted
wave; right: reflected wave. Panel rows: top (a-c) -- upslope propagation,
from 50 m to 1 m; middle (d-f) -- downslope propagation from 50 m
to 100 m; bottom (g-i) -- modal variance normalized by the maximum
value. Comparing panel (f) with (g) and (h): in both upslope and downslope
propagation cases, the reflected energy flux fraction outside the
red box in the lower panels is < 5\%. The red box indicates the spectral
band subjected to strong reflection. \label{fig: da0.5spt}}
\end{figure}

The KdV equation (\ref{eq: tkdv scaled}) is integrated using a symmetric
split step method that combines the time Fourier transform with a
4th order Runge-Kutta integration of the nonlinear term \citep{Sheremet2016}.
In all simulations presented here, the total momentum was conserved
to machine error ($10^{-25}$) and the energy flux was conserved with
an error of $10^{-7}$.

The reflected and transmitted waves are calculated within the framework
of linear theory as if the slope toe were located at a chosen distance
to the position of the initial perturbation; for a given initial pulse
by varying the position of the toe we vary the stage of initial pulse
disintegration. We find the reflection and transmission by applying
the linear theory reflection and transmission coefficients given by
equations \ref{eq: RT up}-\ref{eq: RT down}) to the Fourier transform
of the incoming wave, 
\begin{equation}
\varphi_{R}(t)=\int_{-\infty}^{\infty}\mathcal{R}(f)\hat{\varphi}(f)e^{2\pi ift}df,\;\text{where}\;\hat{\varphi}(f)=\int_{-\infty}^{\infty}\varphi(t)e^{-2\pi ift}dt,\label{eq: Fourier}
\end{equation}
where the subscript $R$ denotes the reflected wave. For the transmitted
component one has to replace $\mathcal{R}$ with $\mathcal{T}$ and
the subscript $R$ with the subscript $T$ . The reflected and transmitted
fractions of the incoming energy flux are 
\begin{equation}
F_{R}=\int_{-\infty}^{\infty}\frac{1}{2}\left|\hat{\varphi}_{R}(f)\right|^{2}df,\;\text{and}\;F_{T}=\sqrt{\frac{h_{2}}{h_{2}}}\int_{-\infty}^{\infty}\frac{1}{2}\left|\hat{\varphi}_{T}(f)\right|^{2}df.\label{eq: flx fractions}
\end{equation}
where $h_{1,2}$ are the two flat bottom depths of the problem (see
figure \ref{fig: problem}a). The formulae are valid both for the
upslope and downslope propagation. The specific expressions for $\mathcal{R}$
with $\mathcal{T}$ are given by \ref{eq: RT up}-\ref{eq: RT down}.

The DSW effect on reflection is determined primarily by two time scales:
the reflection time scale $T_{12}$ (specified by equation \ref{eq: T12})
and the characteristic time scale $T$ of the wave itself. The frequency
band allowed by the reflection and transmission filters (\ref{eq: RT up}-\ref{eq: RT down})
depends on the relation between these two time scales. For sufficiently
gentle or too steep slopes $\alpha$, i.e., $T\ll T_{12}$ or $T\gg T_{12}$,
the contribution of the DSW to transmission is small, because the
initial perturbation is either almost entirely transmitted or nearly
entirely reflected. Therefore, the maximum effect of the DSW is realized
for the waves of the time scales comparable to the reflection time
scale, i.e. $T\approx T_{12}$.

\begin{figure}
\centering
\includegraphics[width=0.8\textwidth]{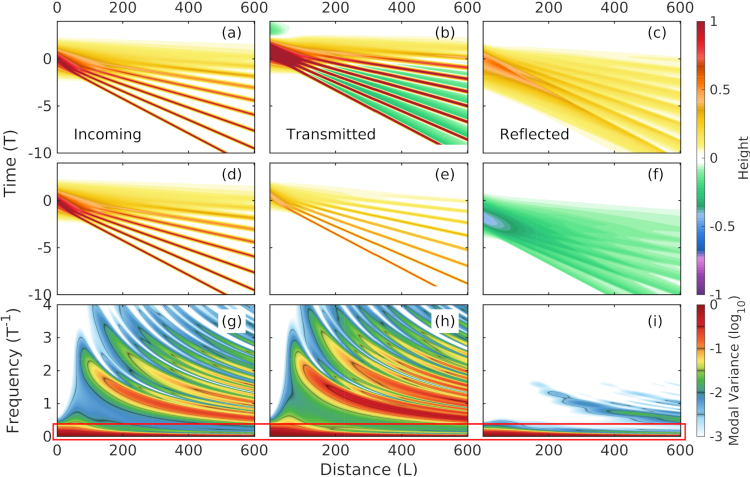}

\caption{Same as in figure \eqref{fig: da0.5spt}, but for a wave with $\sigma^{2}=795$(reference
wave amplitude 1.0 m).\label{fig: da1.0spt}}
\end{figure}

As a reference example in the meteotsunami context consider an initial
single hump elevation of $L\approx3.3$ km with the time scale $T=150$
s starting its propagation either towards the shoreline or in the
opposite direction at a shelf of depth $h=50$ m, which implies $\mu=2.5\times10^{-5}$.
A wave height of $a=1$ m corresponds to $\epsilon=0.02$, and $\sigma^{2}=795$,
In the absence of dispersion such a wave would reach the gradient
catastrophe point after propagating over a distance of $\approx40\ L$,
or $\approx$134 km; from equation \eqref{eq: sol}; its disintegration
is expected to produce at most 12 solitons. The reflected and transmitted
components are calculated below for an upslope transition from 50
m to 1 m depth (similar to the reflection at an inner shelf and beach),
and a downslope transition from 50 m to 1000 m. The 1-m and 1000-m
depths only determine value of the long-wave coefficients (equation
\ref{eq: RT long wave}; i.e., the scale of the $\mathcal{R}(f)$
function), but otherwise have no effect on its shape.

Figure \eqref{fig: RT-slopes} shows reflected fraction of the energy
flux: modulus squared of the reflection coefficient) for different
upslope magnitudes, rescaled in the time scale of the perturbation.
Downslope reflection has a similar dependence on the slope magnitude
(not shown). The maximum effect of the DSW disintegration on reflection
occurs for slopes 0.01$\leq\alpha\leq0.02$ for upslope propagation,
and for 0.05$\leq\alpha\leq0.1$ for downslope propagation.

\subsection{Simulation results\label{subsec: sims}}

The results of numerical simulations presented below correspond to
$\sigma^{2}=397$ and $\sigma^{2}=795$; for the reference wave discussed
above, these values correspond, respectively, to initial wave heights
of 0.5 m and 1.0 m. The simulations were carried out assuming the
slope value $\alpha\approx0.015$ for upslope propagation (blue line
in figure \ref{fig: RT-slopes}), which is a rather typical value
for the USA Atlantic inner shelf. For downslope propagation we set
the slope $\alpha=0.07$.

Because the basis of the DSW evolution effect on reflection is nonlinear
energy transfer away from the strong reflection spectral band indicated
by red rectangle in figures \ref{fig: RT} and \ref{fig: RT-slopes},
before studying DSW evolution effect, it is worth first examining
the effect of the nonlinear steepening considered in isolation. To
this end the nondispersive evolution of the initial perturbation \eqref{eq: sech2}is
shown in figure \eqref{fig: na0.5} for $\sigma^{2}=397$ (reference
wave initial amplitude $a=0.5$ m). The evolution of the shapes of
incoming, reflected and transmitted waves are shown in figure (\ref{fig: na0.5},
upper panels). The incoming wave exhibits the characteristic evolution
of the Riemann simple wave \citep[e.g.,][]{Whitham1973}: the height
is constant, the front steepens and eventually becomes vertical (gradient
catastrophe). The distance to the gradient catastrophe point is $\approx80\ L$
(266 km for the reference wave). The height of the reflected and transmitted
waves are 0.58, and 2.38 (the latter taking into account shoaling).
During the reflection/transmission process the total energy flux is
conserved; as the incoming wave steepens, however, bound Fourier components
appear in the energy flux spectra at frequencies outside the strong
reflection band indicated by red boxes in figure \ref{fig: na0.5},
lower panels. For these modes, the fraction of the transmitted energy
flux increases significantly. Still, the variance of these bound modes
is bounded from above by an $f^{-2}$ decay (characteristic of e.g.,
square pulse variance spectrum), therefore the effect is limited to
a $\approx3.5$ \% increase in the transmitted flux fraction.

\begin{figure}
\centering
\includegraphics[width=0.8\textwidth]{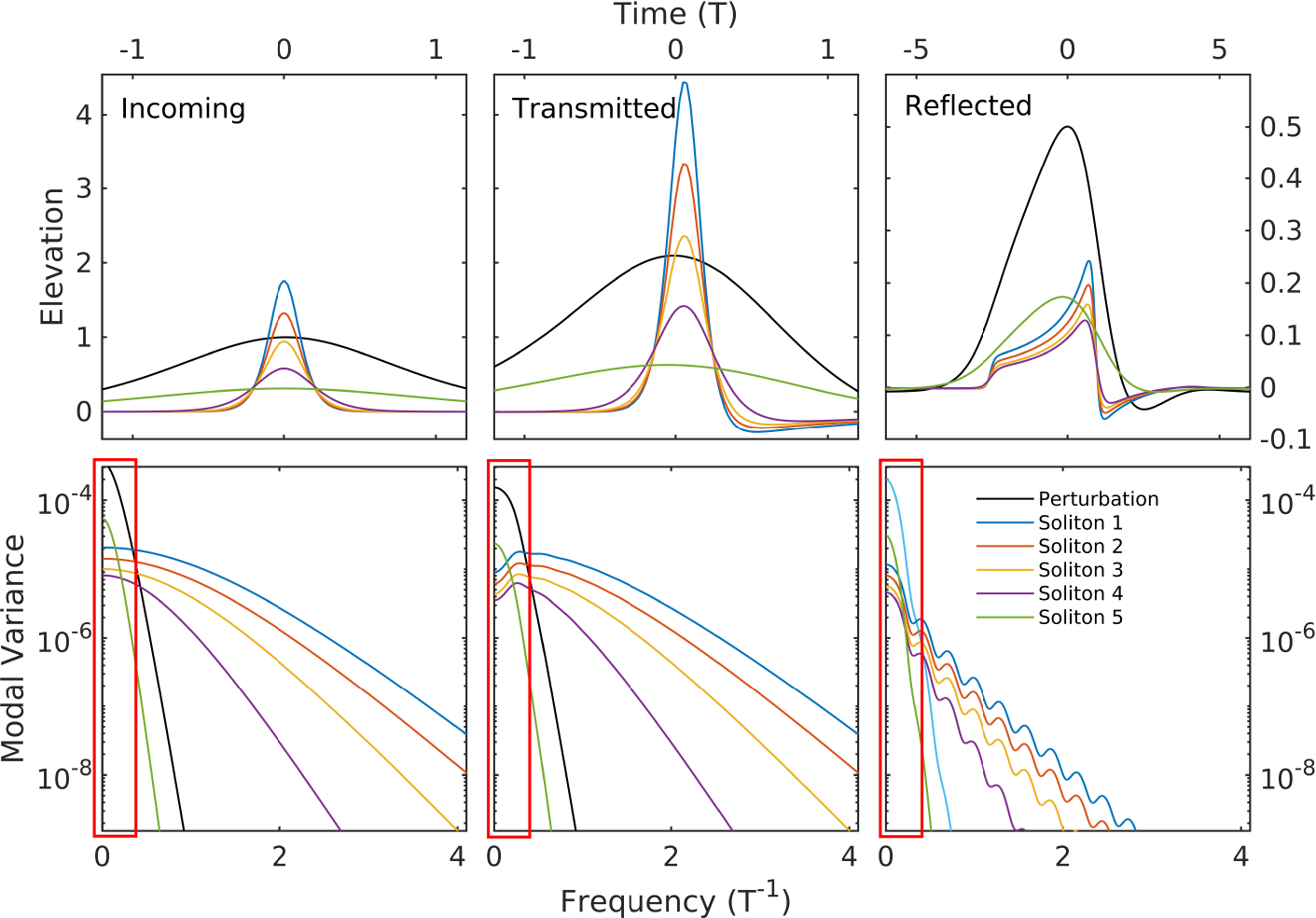}

\caption{Solitons identified in the numerical simulation of dispersive evolution
of perturbation with $\sigma^{2}=397$ (reference wave amplitude 0.5
m). Upper panels: free surface elevation. Lower panels: modal variance.
Panel columns: left -- incoming wave; middle: transmitted wave; right:
reflected wave. The reflected energy flux fraction outside the red
box in the lower panels is less than 10\%. \label{fig: da0.5sol}}
\end{figure}

The effect of the DSW evolution on reflection/transmission is much
stronger than the effect of nondispersive evolution. The incoming,
reflected and transmitted waves during nondispersive evolution are
shown in figure (\ref{fig: na0.5}) for $\sigma^{2}=397$ $a=0.5$
m. The incoming wave exhibits the characteristic evolution of the
simple wave: the height is constant, the front steepens and eventually
becomes vertical (gradient catastrophe). The distance to the gradient
catastrophe point is $80L=240$ km. The height of the reflected and
transmitted waves are 0.58 m, and 2.38 m. During the reflection/transmission
process the total energy flux is conserved, but as the incoming wave
steepens, bound Fourier components appear in the energy flux spectra
at higher frequency. In these frequency bands (compare with figure
\ref{fig: RT}), the fraction of the reflected energy flux decreases,
while the transmitted fraction increases. Overall, the effect of the
bound wave components, is relatively weak, with the maximum increase
of transmitted fluxes reaching approximately 3.5 \%. In contrast,
the spectral energy transfers associated with the formation of the
solibore and its subsequent disintegration into a train of solitons
have much stronger effects on the reflection and transmission of long
waves. We illustrate these effects for the perturbation \eqref{eq: kdv IC},
\eqref{eq: sech2} for Ursell numbers $\sigma^{2}=397$ and $\sigma^{2}=795$;
for the reference wave, these values correspond to amplitudes $a=0.5$
m and $a=1.0$ m, and nonlinearity parameters $\epsilon=0.01$ and
$\epsilon=0.02$; the dispersion parameter is the same for both waves,
$\mu=2.5\ 10^{-5}$. Based on equations \eqref{eq: sol} the disintegration
should produce up to 8 and up to 12 solitons, respectively.

\begin{figure}\centering
\includegraphics[width=0.6\textwidth]{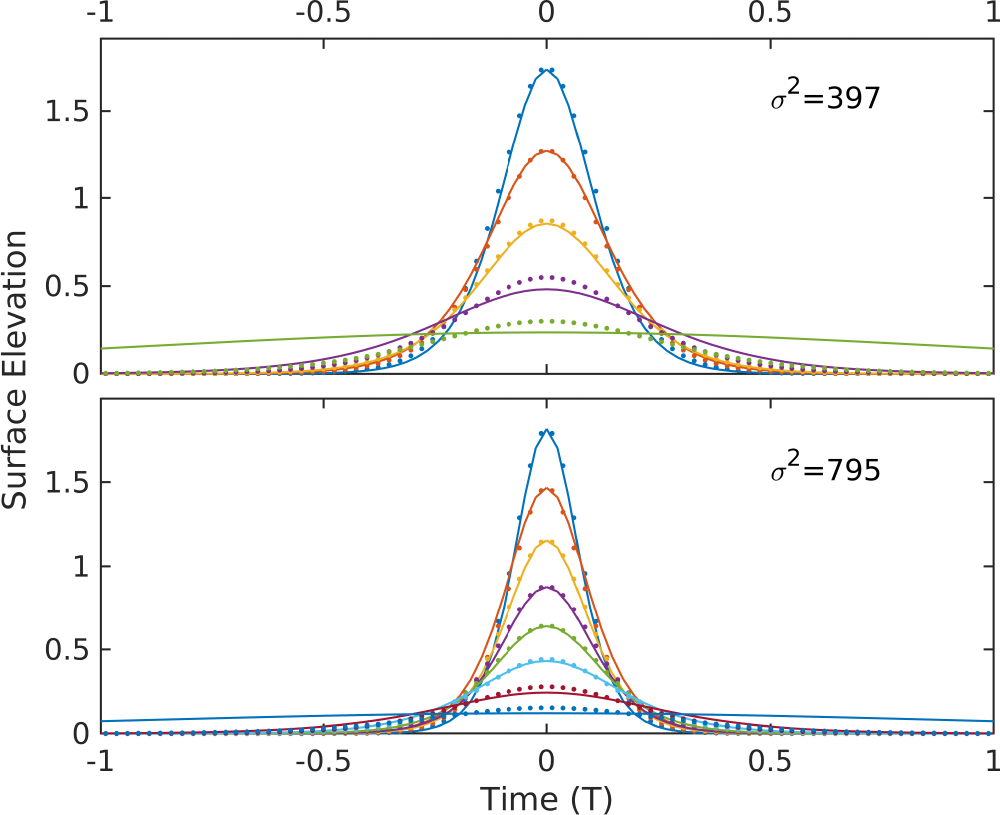}

\caption{Comparison of solitons identified at the last station (propagation
distance $\approx600\ L$, or 2,000 km for the reference wave) in
the numerical simulation (continuous lines) of the evolution of the
perturbation with $\sigma^{2}=397$ (figure \ref{fig: da0.5spt})
with the corresponding asymptotic solution (dots; equation \ref{eq: sol}).
In this case the DSW evolution is expected to produce up to 8 solitons,
but at the chosen distance only 5 may be reliably identified in the
simulation. While the taller solitons are nearly identical to the
asymptotic solution, the weaker ones are not fully formed yet.\label{fig: NA-sol}}
\end{figure}

The simulations (figures \ref{fig: da0.5spt}-\ref{fig: da0.5sol})
illustrate the importance of intermediate stages of the DSW evolution.
Because the nonlinearity is relatively strong, in the sense that $\sigma^{2}\gg\sigma_{s}^{2}=12$,
in both cases the wave at first evolves as a nondispersive Riemann
wave for some considerable distance, $\approx30$ $L$, ($\approx100$
km, reference wave) for the higher wave, twice that for the weaker
wave. Despite the long integration domain, $\approx1,000\ L$ ( $\approx$
3,000 km, reference wave), the perturbation does not reach the asymptotic
state of the full soliton separation assumed by equations \eqref{eq: sol}.
Equations \eqref{eq: sol} predict up to 8 solitons for $\sigma^{2}=397$,
but only 5 solitons can be distinguished as truly separate, with perhaps
the last two just beginning to form toward the end of the domain (figure
\ref{fig: da0.5spt}, upper panels). Similarly, of the up to 12 solitons
predicted for $\sigma^{2}=795$ only 8 can be confidently identified
in figure \ref{fig: da1.0spt}, upper panels. 
\begin{figure}\centering
\includegraphics[width=0.6\textwidth]{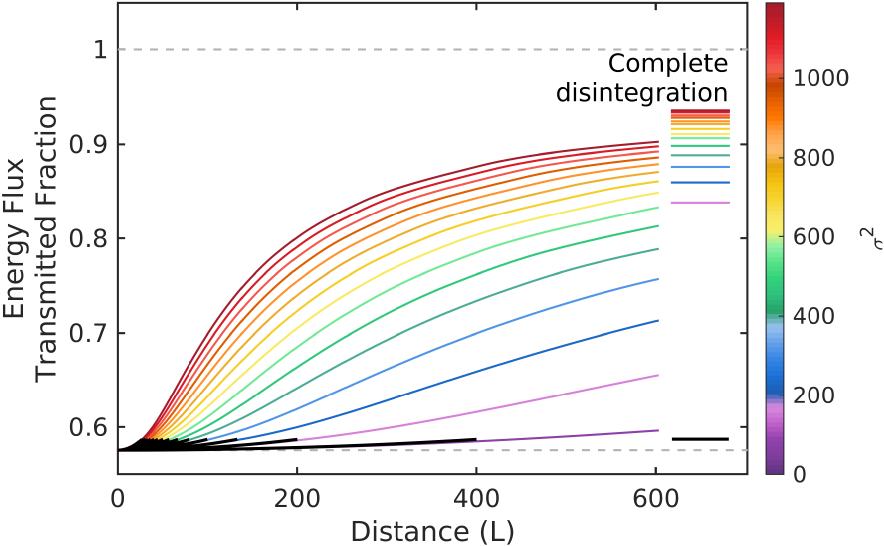}

\caption{Transmitted fraction of total energy flux for initial perturbation
with $80\le\sigma^{2}\le1,200$, or, for the reference wave with the
amplitudes $a$ in the range:, $0.1\le a\le1.5$ m. Black lines represent
the transmitted fraction for nondispersive propagation. The lines
in the detached ``column'' to the right show the transmitted flux
calculated employing the asymptotic solution \eqref{eq: sol}. The
total propagation distance is $\approx600\ L$, or 2,000 km for the
reference wave. \label{fig: allflx}}
\end{figure}

The effect of the DSW disintegration on the nonlinear energy transfer
away from the strong reflection band is most evident in figures \ref{fig: da0.5spt}-\ref{fig: da1.0spt},
lower panels: modal variance increases significantly at high frequencies,
i.e. outside the red box at high transmission band, as soon as solitons
form. At the end of the integration domain, the soliton contributions
dominate the transmitted wave. The key feature is the total variance
outside the strong reflection band indicated by red box; the variance
modulation (shifting lobe appearance) is only the result of estimating
modal variance using the Fourier transform of the entire time series.
Solitons emerge ordered by their height, the tallest ones move faster.
The Fourier estimate converts soliton separation into phase modulation.
The modulation disappears if soliton components are considered in
isolation (figure \ref{fig: da0.5sol}, lower panels). In numerical
simulations, however, soliton identification is possible only for
well separated leading solitons (figure \ref{fig: da0.5sol}, upper
panels).

Figure \ref{fig: da0.5sol} also details characteristics of the DSW
disintegration that affect reflection. Because the Ursell number of
a soliton is fixed, under adopted scaling $\sigma^{2}=12$, taller
solitons are narrower. This implies that the first emerging solitons
are more effective in redistributing energy flux to higher wavenumber
modes with high transmission rates. The time scale of lagging is comparable
with the time scale of the initial perturbation, and show a similarly
strong reflection.

Increasing the nonlinearity of the initial perturbation specified
by $\sigma^{2}$ affects reflection in several ways: (i) the DSW disintegration
occurs earlier (compare figures \ref{fig: da0.5spt}-\ref{fig: da1.0spt},
upper panels); (ii) the number of emerging solitons increases (equation
\ref{eq: sol}; figures \ref{fig: da0.5spt}-\ref{fig: da1.0spt});
(iii) the height of resulting solitons increases (equation \ref{eq: sol}),
which means that the energy carried by each soliton increases (equation
\ref{eq: solQ}), and the emerging solitons are narrower (figure \ref{fig: NA-sol});
(iv) the reflected energy flux decreases.

The DSW evolution may be also described as a disintegration into ``particles''
(solitons) that carry each a ``quantum'' of energy flux (equation
\ref{eq: solQ}). Smaller scale particles carry higher quantum of
energy flux and have higher transmission rates. While the nature of
the reflection coefficient makes a slope opaque to long waves, the
DSW disintegration creates small scale quanta for which the slope
is transparent. This effect increases with the nonlinearity of the
initial perturbation. Note that the solitons of small amplitude might
be too broad and, therefore, experience strong reflection.

In applications, the speed of evolution is essential, because the
disintegration process rarely has enough propagation space to form
a complete soliton sequence predicted by theory at large times. Figure
\eqref{fig: NA-sol} compares the solitons identified in the solution
after propagating for $\approx450\ L$ (reference wave, $\approx1,500$
km), with the asymptotic solution (equation \ref{eq: sol}; \citealp{Karpman1967,Karpman1975}).
It is clear that while the tall solitons are identical to the asymptotic
form, the weaker solitons are not fully formed yet. As stated above,
this implies that the stage of the DSW disintegration is important
for oceanographic applications, where the full, asymptotic state of
the disintegration may not be observable. This is also illustrated
in figure \eqref{fig: allflx}, which summarizes the evolution of
the transmitted fraction of the total incoming energy flux for $80\leq\sigma^{2}\leq1,200$
(reference wave: $0.1\leq a\leq1.5$ m) for a distance of propagation
of $\approx450\ L$ (reference wave, $\approx2,000$ km). The transmitted
fraction of total energy flux increases with the initial height or
$\sigma^{2}$ and the degree of soliton separation. Over a flat bottom
the process approaches saturation as the solitons approach the asymptotic
state, but the distance for this may be unrealistic for application
to long wave propagation over the continental shelf. As noted before,
the simple wave deformation prior to the gradient catastrophe accounts
for only $\approx3$\% increase in transmission efficiency.

\section{Discussion\label{sec: discussion}}

Here, we discuss the robustness of the results and their sensitivity
to the underlying assumptions of the Boussinesq equations and maximally
simplified bathymetry models used here. The main result of this study
is the deconstruction of self-induced transparency of long waves propagating
over a localized inhomogeneity of bathymetry, into nonlinearity, dispersion,
and reflection elements. 

Within the parameter domain studied, the DSW disintegration is effective
in transferring energy flux across the scale boundary that separates
low reflection. As a result, the self-induced transparency is an order
one effect: reflection drops from order one to nearly zero. The mechanism
works both for up- and downslope bathymetric inhomogeneity. The generation
of bound harmonics, dominant for non-dispersive waves, has a much
weaker effect (up to 3.5\%) on reflection/transmission. Whether this
conclusion holds for different bathymetries requires further investigation.

For understanding the DSW dynamics, the flat bathymetry simplification
has the advantage that the long term asymptotics are known. This allows
for some simple \emph{a priori }estimates of effects of the DSW evolution
on reflection and transmission. For example, a given initial pulse
tends to disintegrate into the known number of solitons with known
amplitudes as in example \eqref{eq: sech2}. The strength of the total
effect is primarily controlled by the magnitude of the Ursell number
$\sigma$. As $\sigma$ increases, the DSW disintegration produces
more, narrower solitons that are more effectively transmitted. The
smallest soliton is also the widest one; if the smallest soliton is
still narrow enough to be much narrower then the length of the slope,
then all of them can pass through effectively without reflection.
This provides an easy way to calculate the upper bound of the transmitted
energy flux for the asymptotic state of fully separated solitons.
The few examples in \S\ref{sec: results} suggest, however, that
for realistic bathymetries only the first few solitons are likely
to get fully separated, which considerably overestimates the transmission.
A more accurate estimate could be possible following the recently
found explicit solution accurately capturing DSW evolution over a
flat bottom \citep{Opanasenko2023}.

A key element of the process is the high-pass filter behavior of linear
scattering at a localized inhomogeneity. Although analytical results
supporting the high-pass filter role of bathymetry inhomogeneity exist
only for a few geometrically simple inhomogeneity models \citep{Meyer1979,Meyer1975,Mei2005,ErmakovStepanyants2020},
the behavior of small scale harmonics can be always described employing
the WKB approximation and, hence, with negligible reflection, which
suggests that this behavior is universal within the framework of linear
theory. One can argue that it is also valid for weakly nonlinear waves.
Indeed, for a real bathymetry profile, one can always introduce an
inhomogeneity scale $\Lambda$, playing the same role as the slope
length $L_{12}$, see figure (\ref{fig: problem}b). Estimating the
nonlinear spatial scale (distance to gradient catastrophe) as $\frac{L}{\epsilon}$,
the assumption that $\Lambda\ll\frac{L}{\epsilon}$ allows us to neglect
nonlinearity on the slope, greatly simplifying the description. All
our results are obtained for this specific regime, which is applicable
to a wide range of realistic situations. Even when this assumption
is not valid, the self-induced transparency phenomenon does not disappear,
it just becomes more complex and exhibits new dynamics. For example,
for high enough initial nonlinearity, pulses resulting from the ``primary''
DSW disintegration could develop ``secondary'' DSWs, further enhancing
the phenomenon; or could become strongly nonlinear and break. Both
scenarios are interesting and therefore merit dedicated studies.

Replacing the highly simplified flat bathymetry used here with sufficiently
mild slopes ensuring negligible reflection instead of the flat shelves
cannot disrupt the effectiveness of DSW disintegration, and therefore
preserves the DSW self-induced transparency. Qualitatively, for shoaling
waves the number of solitons produced increases while their scale
decreases, which has the effect that the transmitted fraction of the
energy flux increases steadily instead of saturating as for the flat
bathymetry case. This process could be studied using the variable
coefficient KdV equation without significant additional numerical
effort. 

In this study, the nonlinear interaction between incoming and reflected
waves was neglected. This approximation is justified if nonlinearity
is small, which implies that interaction time counter-propagating
wave pulses is too small for nonlinear effects to accumulate, even
when we take into account the spreading of the incoming and reflected
wave pulses. The nonlinear interaction between two counter-propagating
pulses was thoroughly studied by \citet{Khusnutdinova2012,Khusnutdinova2014}
from a different perspective. We are not aware of dedicated studies
of nonlinear interaction of two counter-propagating waves in the Boussinesq
type equations with inhomogeneity, but we expect that the same logic
as in the homogeneous case should be applicable. This assumption could
be verified by direct integration of the Boussinesq equations.

These arguments suggest that the key elements of the mechanism, the
inhomogeneity high-pass filter and DSW disintegration, are robust and
not sensitive to tweaking either the model or topography. Moreover,
even if we take into consideration the factors and effects \emph{a
priori} neglected, such as three-dimensional bathymetry and wave fields,
bottom friction, and interaction with atmosphere, we do not see a
candidate mechanism potentially able to destroy the phenomenon. 

We conclude, by noting that self-induced transparency involves robust
physical elements, therefore the process itself must be widely robust.
Boussinesq type equations with inhomogeneity play fundamental role
in many physical contexts, such as e.g. long internal gravity waves
\citep[e.g.,][]{Grimshaw1998}, plasmas \citep{Karpman1975}, nonlinear
waves in solids \citep{Khusnutdinova2023}. Scattering at localized
homogeneities is a universal phenomenon, and its high-pass filter behavior
is universal and well documented in the framework of linear theory
(e.g., \citealp{Felsen1994,Lekner2016}). Furthermore, we expect the
phenomenon to be more general than the Boussinesq type equations with
inhomogeneity and be applicable to a wide class of weakly dispersive
systems. These expectations remain to be properly developed and verified.

Although the calculations presented in the paper illustrate relevance
of the self-induced transparency for long ocean waves, the results
are too simplified to be directly applied to real-life tusunami/meteotsunami
events. Quantifying the DSW disintegration role for any specific conditions
requires extensive and expensive direct integration of the Boussinesq
equations for the specific bathymetry and a range of parameters of
incoming waves. For long ocean waves, numerical models are readily
accessible (e.g., FUNWAVE-TVD \citealp{Shi2016}, tested and validated
over many years), but this task goes beyond the scope of present work.



\bibliographystyle{plainnat}
\bibliography{SheremetShrira-ams}


\end{document}